\documentclass[aps,nofootinbib,notitlepage,superscriptaddress,twocolumn,10pt,prd]{revtex4-1}

\usepackage{graphicx}
\usepackage{amsmath,amssymb}
\usepackage{hyperref}
\usepackage{braket}
\usepackage{subfigure}
\usepackage{float}
\usepackage[dvipsnames]{xcolor}
\usepackage{mathrsfs}
\usepackage{comment}
\usepackage{multirow}
\usepackage{soul}

\def\H{\mathcal{H}}
\def\DT{\Delta_\varphi}

\begin{document}

\title{Detecting Dark Energy Fluctuations with Gravitational Waves}

\author{Alice Garoffolo}
\affiliation{Institute Lorentz, Leiden University, PO Box 9506, Leiden 2300 RA, The Netherlands}
\author{Marco Raveri}
\affiliation{Center for Particle Cosmology, Department of Physics and Astronomy, University of Pennsylvania, Philadelphia, PA 19104, USA}
\author{Alessandra Silvestri}
\affiliation{Institute Lorentz, Leiden University, PO Box 9506, Leiden 2300 RA, The Netherlands}
\author{Gianmassimo Tasinato}
\affiliation{Department of Physics, Swansea University, Swansea, SA2 8PP, U.K.}
\author{Carmelita Carbone}
\affiliation{Istituto di Astrofisica Spaziale e Fisica cosmica Milano, Via A. Corti 12, I-20133 Milano, Italy}
\affiliation{INFN Sezione di Milano, Via G. Celoria 16, I-20133, Milano, Italy}
\author{\\Daniele Bertacca}
\affiliation{Dipartimento di Fisica e Astronomia Galileo Galilei, Universit\`{a} di Padova,  I-35131 Padova, Italy}
\affiliation{INFN Sezione di Padova,  I-35131 Padova, Italy}
\author{Sabino Matarrese}
\affiliation{Dipartimento di Fisica e Astronomia Galileo Galilei, Universit\`{a} di Padova,  I-35131 Padova, Italy}
\affiliation{INFN Sezione di Padova,  I-35131 Padova, Italy}
\affiliation{INAF - Osservatorio Astronomico di Padova, Vicolo dell Osservatorio 5, I-35122 Padova, Italy}
\affiliation{Gran Sasso Science Institute, Viale F. Crispi 7, I-67100 L'Aquila, Italy}

\begin{abstract}
Luminosity distance estimates from electromagnetic and gravitational wave sources are generally different in models of dynamical dark energy and gravity beyond the standard cosmological scenario.
We show that this leaves a unique imprint on the angular power-spectrum of fluctuations of the luminosity distance of gravitational-wave observations which tracks inhomogeneities in the dark energy field.
Exploiting the synergy in supernovae and gravitational wave distance measurements,
we
%it is possible to
 build a joint estimator that directly probes dark energy fluctuations, providing a conclusive evidence for their existence in case of detection.
Moreover, such measurement would also allow to probe the running of the Planck mass.
We discuss experimental requirements to detect these signals.
\end{abstract}

\maketitle
\section{Introduction}
Over the last decades, a variety of cosmological data have confirmed $\Lambda$CDM as the standard model of cosmology~\cite{Aghanim:2018eyx,Abbott:2017wau}.
Despite its successes, the physical nature of its main components still eludes us. In particular, understanding whether cosmic acceleration is sourced by a cosmological constant, $\Lambda$,
or rather by dynamical dark energy (DE) or modifications of the laws of gravity (MG)  is  one of the main science drivers of upcoming cosmological missions. In the presence of DE/MG, the dynamical degrees of freedom of the theory change, generally with the appearance of a new scalar field to which we broadly refer as the  ``DE field''.  The latter, leaves imprints not only on the dynamics of the Universe, but also on the clustering and growth of large-scale cosmological structures. Next generation galaxy surveys~\cite{Amendola:2016saw, Bacon:2018dui, Dore:2014cca,Ivezic:2008fe} aim at constraining these effects and, possibly, {\it indirectly} detecting DE/MG.

The  detection of gravitational waves (GW) has opened a new observational window onto our Universe, promising to offer complementary probes to shed light on cosmic expansion.
GW events at cosmological distances can be used as {\it standard sirens}~\cite{Schutz:1986gp,Holz:2005df,Cutler:2009qv} for measuring the expansion rate of the universe.
This recent approach is complementary to measuring the luminosity distance of {\it standard candles}, like Type-Ia supernovae (SN).
Multi-messenger observations can also be used to test theories of modified gravity, as recently reviewed in~\cite{Ezquiaga:2018btd}.

On the homogeneous and isotropic background, luminosity distances depend only on redshift, leading to the standard distance-redshift relation. Inhomogeneities in the Universe  induce a dependence of the distances on direction. Fluctuations in the  EM luminosity distance constitute an important probe for cosmology and have been well studied~\cite{Sasaki:1987ad,Pyne:2003bn,Holz:2004xx,Hui:2005nm,Bonvin:2005ps}, while the case of GWs has been addressed in General Relativity (GR) in~\cite{Laguna:2009re,Kocsis:2005vv,Bonvin:2016qxr,Hirata:2010ba,Dai:2016igl,Mukherjee:2019wfw,Mukherjee:2019wcg}.
In presence of DE/MG,  the GW luminosity distance generally differs from the one traced by electromagnetic (EM) signals, both at the  unperturbed, background level~\cite{Belgacem:2017ihm,Belgacem:2018lbp} and in its large-scale fluctuations~\cite{Garoffolo:2019mna,Dalang:2019rke}. Importantly,  fluctuations in the EM luminosity distance are affected by the DE field only indirectly while, as first shown in~\cite{Garoffolo:2019mna}, linearized fluctuations of the GW luminosity distance contain contributions directly proportional to the clustering of the DE field.

\begin{figure*}[t]
\centering
\includegraphics[width=\textwidth]{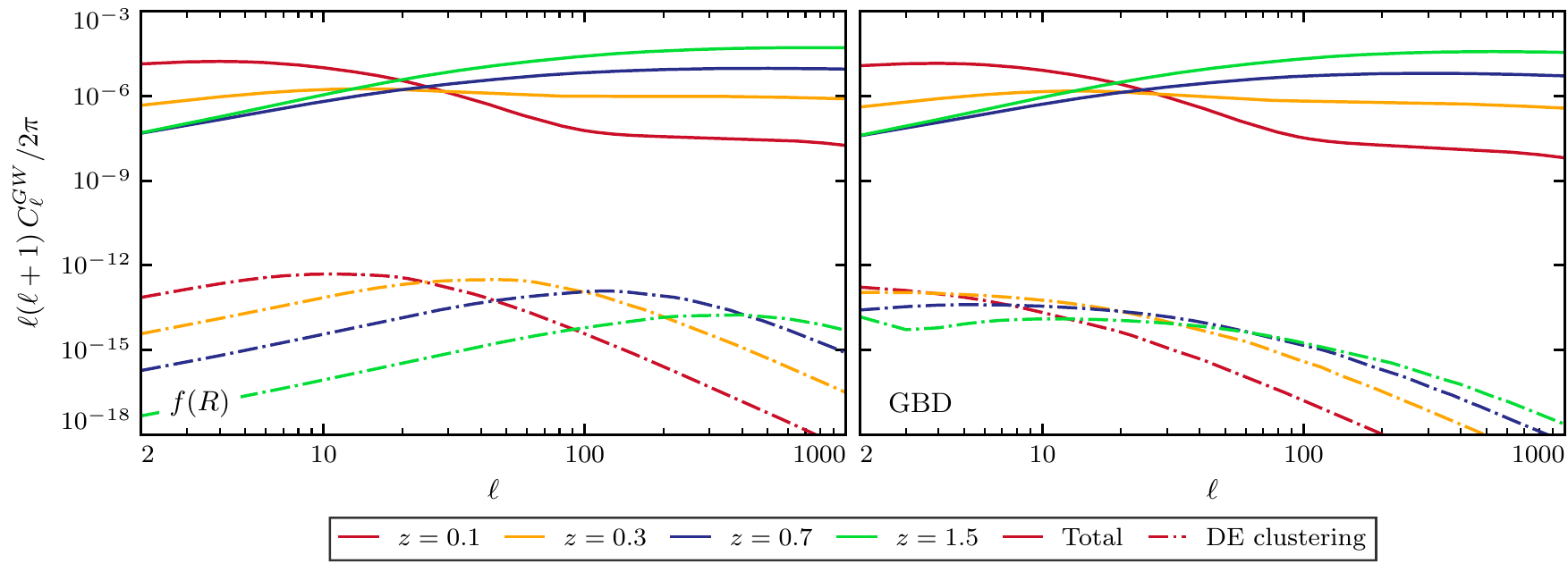}
\caption{
\label{Fig:PowerSpectrum}
Angular power-spectrum of gravitational-wave luminosity distance fluctuations. Solid lines show the total power-spectrum, dashed lines the scalar field clustering component.
}
\end{figure*}

In this work we combine SN and GW luminosity distance fluctuations into a novel estimator to {\it directly} detect the  signal of DE clustering.
This signal can not be mimicked by other effects and would provide convincing evidence for the existence of the DE field.
If DE does not directly couple to known particles through non-gravitational interactions, ours
is a promising method to pursue its direct detection.
The approach we propose allows  to probe the DE field at cosmological scales, far from sources that can hide its presence by means of screening mechanisms (see e.g.~\cite{Joyce:2014kja,Babichev:2013usa,Burrage:2017qrf}).

\section{The GW luminosity distance power-spectrum in DE/MG}
The luminosity distance, as inferred by an EM or GW signal propagating through a universe with structures, depends on the  observed redshift, $z$, and on the direction of arrival in the sky, $\hat{\theta}$.
We decompose the observed luminosity distance of a source as a sum of its background and fluctuation components, i.e. $d_L(\hat{\theta}, z)=\bar{d}_L(z)+\Delta d_L(\hat{\theta},z)$.
We work in the context of \emph{scalar-tensor} theories of gravity, which encompass most of the candidate DE/MG models. These theories are characterized by  a non-minimal coupling of the DE field to space-time curvature. This causes a running of the Planck mass, $M_P$, which generally depends on the background configuration of the DE field $\varphi$, and on its first derivatives, through $X = - \nabla_\mu \varphi \nabla^\mu \varphi /2$.
More specifically, we consider DHOST theories~\cite{Langlois:2015cwa,Crisostomi:2016czh,Langlois:2017mxy} (see e.g.~\cite{Langlois:2018dxi}), focusing on scenarios
%a set-ups
 that ensure luminal speed for GWs and avoid instabilities associated with graviton decay into DE~(\cite{Crisostomi:2019yfo}). We require that high-frequency scalar fluctuations propagate at the same speed of tensor modes, as discussed in~\cite{Garoffolo:2019mna}.
The dependence of $M_P$ on the DE field gives new contributions to $d_L^{\rm GW}$ with respect to the GR case.
At the background level one finds $\bar d^{\rm GW}_L = [ M_P(z) / M_P(0) ]\bar d_L$, where $\bar d_L=(1+z)\int_0^z d\tilde{z}[(1+\tilde{z})\H]^{-1}$ is the luminosity distance associated to electromagnetic sources, with $\mathcal{H} = a'/a$ the Hubble parameter.
The multiplicative factor $M_P(z) / M_P(0)$ accounts for the extra friction acting on the GWs during their propagation induced by the running of $M_P$.

At the linear level in fluctuations, generalizing to DHOST the procedure of~\cite{Garoffolo:2019mna} as shown in Appendix~\ref{App1}, we find:
\begin{align} \label{Eq:DeltaGW}
\frac{\Delta d^{\rm GW}_L}{\bar{d}^{\rm GW}_L} =&  - \kappa - (\phi+\psi)+\frac{1}{{\chi}}\int^{{\chi}}_0 d\Tilde{\chi}\,(\phi + \psi)  \nonumber\\
	&  +\phi \bigg(\frac{1}{\H{\chi}} -\frac{M'_{P}}{\H M_P}\bigg)+ v_{\|} \bigg(1-\frac{1}{\H{\chi}} +\frac{M'_{P}}{\H M_P} \bigg)   \nonumber\\
	&  - \left(\,1  - \frac{1}{\H{\chi}} + \frac{M'_{P}}{\H M_P}\right) \int^{{\chi}}_0 d\Tilde{\chi}\,(\phi' + \psi') \nonumber\\
	&  + \frac{ M_{P, \varphi}}{M_P} \delta \varphi + \frac{ M_{P, X}}{M_P} \delta X \,.
\end{align}
where a prime indicates differentiation w.r.t. conformal time,
$\kappa$ denotes the weak lensing convergence,
$\chi$  the comoving distance to the source,
$\phi$ the Newtonian potential,
$\psi$ the intrinsic spatial curvature potential, and
$v_{\|}$ the component along the line of sight of the peculiar velocity of the source:
all in Poisson gauge and following the conventions of~\cite{Garoffolo:2019mna}.
$M_{P, \varphi}$ and $ M_{P, X}$ stand for the derivative of $M_P(\varphi, X)$ w.r.t. its arguments.

The physical effects contributing to $\Delta d^{\rm GW}_L$ are: lensing convergence, volume dilation, and time delay in the first line of Eq.~\eqref{Eq:DeltaGW}, that are only indirectly influenced by DE/MG; Sachs-Wolfe (SW), Doppler shifts, and Integrated Sachs-Wolfe (ISW) effects in the second and third lines, which show an additional explicit decay that depends on the evolution of $M_P$; damping due to DE field inhomogeneities in the fourth line of Eq.~\eqref{Eq:DeltaGW}. These last effects are the main interest of this work as they are {\it unique} to the GW luminosity distance fluctuations.

We use Eq.~\eqref{Eq:DeltaGW} to build the angular power-spectrum of GW luminosity distance fluctuations averaged over a given redshift distribution of the sources
\begin{equation}\label{C_ell_GW}
C^{\rm GW}_\ell=4\pi\int \text d \ln k  \left(\frac{\Delta d^{\rm GW}_L}{\bar{d}^{\rm GW}_L}\right)^{W}_{k \, \ell}\left(\frac{\Delta d^{\rm GW}_L}{\bar{d}^{\rm GW}_L}\right)^{W}_{k \, \ell} \,,
\end{equation}
where we work in Fourier space for the perturbations, with $k$ being the momentum, and
\begin{equation}
\left(\frac{\Delta d^{\rm GW}_L}{\bar{d}^{\rm GW}_L}\right)^{W}_{k\, \ell} = \int^{\infty}_0 dz \, j_\ell(k \chi) \,W(z) \, \left(\frac{\Delta d^{\rm GW}_L}{\bar{d}^{\rm GW}_L} \right)\,,
\end{equation}
and $j_\ell(x)$ is the spherical Bessel function and $W(z)$ is the source window function, normalized to $1$. The effect of each term in Eq.~\eqref{Eq:DeltaGW} on the angular power-spectrum can be studied independently. We report their explicit forms in Appendix \ref{App3}.
We use the notation $\bar{d}_L(z)$ for the background luminosity distance, to indicate its angular average, weighted by the given redshift distribution.

\begin{figure*}[t]
\centering
\includegraphics[width= \textwidth]{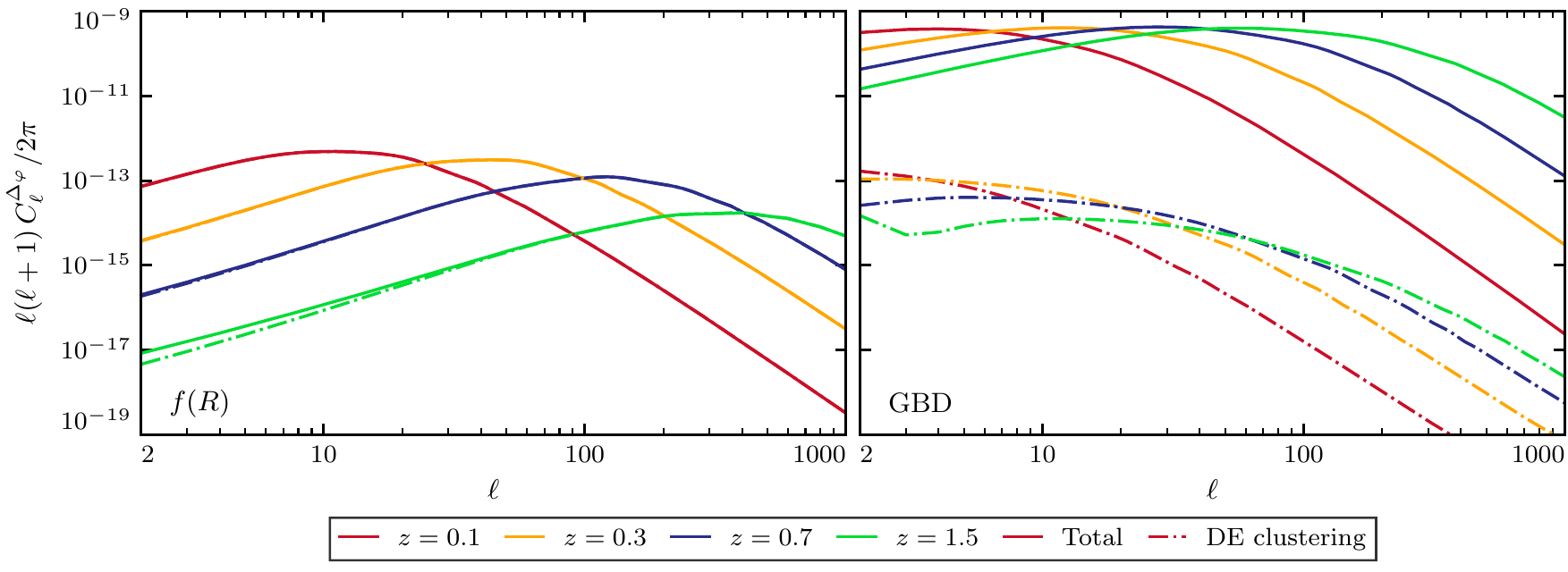}
\caption{
\label{Fig:PowerSpectrumDelta}
The angular power-spectrum of the difference between GW and SN luminosity distance fluctuations.
Solid lines show the total power-spectrum, dashed lines the scalar field clustering component.
}
\end{figure*}

We have implemented the calculation of $C_\ell^{\rm GW}$ in \texttt{EFTCAMB}~\cite{Hu:2014oga},  allowing us to study this quantity for a broad host of DE/MG models. In order to explore  in detail the impact of MG on $C_\ell^{\rm GW}$ we focus for a moment on two representative models. First, a designer $f(R)$ model on a $\Lambda$CDM background~\cite{Song:2006ej}, with the only model parameter set to $B_0 = 10^{-4}$ which is compatible with current constraints~\cite{Peirone:2016wca}. Second, an agnostic parametrization of $M_P$,  such that the ratio $(M'_P/ M_P)$ is a linear function of the scale-factor, $a(z)$, $M'_P / M_P \equiv  (M'_P/ M_P )|_{o} \, a $, where $(M'_P/ M_P )|_{o}$ is the value of the ratio today, which we set to $0.05$.  This minimal parametrization, implemented on a $\Lambda$CDM background, is representative of the Generalized Brans-Dicke (GBD)~\cite{DeFelice:2010jn, Perrotta:1999am, Baccigalupi:2000je} family.
In both these models,  the Planck mass $M_P$ depends on the scalar field value alone. We defer to future works the investigation of
cases where  $M_P$ depends also on $X$.

Figure~\ref{Fig:PowerSpectrum}  shows the angular power-spectrum, $C_\ell^{\rm GW}$, for the two scenarios
described above.
To highlight redshift dependencies, we choose a Gaussian distribution for the GW sources centred in various redshifts $z_i$,  with width $\Delta z = 0.01$, i.e. $W(z)= {\cal N} \exp [- (z -z_i)^2/ (2 \Delta z^2)]$ where ${\cal N}$ is the normalization constant.
The total signal significantly changes shape with increasing redshift. At low redshifts and large scales, the signal is dominated by the Doppler effect, due to the bulk-flow of the environment in which the GW sources are embedded. The Doppler contribution then decays for growing $\ell$, and the angular power-spectrum at small scales is dominated by lensing convergence;
the Doppler term also decays in redshift, while lensing grows and eventually dominates the high-redshift part of the signal.
For both  models considered, the relative behaviour between Doppler and lensing convergence is qualitatively unaltered with respect to GR~\cite{Bertacca:2017vod}.
Figure~\ref{Fig:PowerSpectrum} also shows the direct contribution of $\delta \varphi$ to the total signal, i.e. the last line in Eq.~\eqref{Eq:DeltaGW}. This is of the same order of magnitude in both scenarios, and  results largely subdominant compared to the total signal.
For the $f(R)$ model  the scalar field contribution has a noticeable scale-dependent feature that evolves in time as the Compton wavelength of the model.
At higher redshift,  the Compton scale of the scalar field is smaller and, correspondingly, the feature in the power-spectrum moves to smaller scales.
In the GBD case, on the other hand, any feature in the shape of the power-spectrum is less pronounced, as it only leads to the decay of DE fluctuations below the horizon.

\section{The joint SN/GW estimator}
The direct contributions of DE fluctuations to $C_\ell^{\rm GW}$ is very small compared to other effects, making it impossible to detect their presence in the angular correlations using  GW data only. Interestingly, since photons are not affected directly by DE or MG, $\Delta d^{\rm SN}_L$ is structurally unchanged w.r.t. GR, hence is obtained by neglecting all the explicit DE/MG terms present in Eq.~(\ref{Eq:DeltaGW}) as shown in Appendix \ref{App2}. We can then single out the distinctive DE field contributions, by combining standard sirens and standard candles;  assuming that  we have measurements of both SN and GW at the same redshifts and positions and subtract the two as:
\begin{align} \label{Eq:DeltaEstimator}
\DT(\hat{\theta}, z) \equiv \frac{\Delta d_L^{\rm SN}(\hat{\theta},z)}{\bar{d}_L^{\, \rm SN}}-\frac{\Delta d_L^{\rm GW}(\hat{\theta},z)}{\bar{d}_L^{\, \rm GW}} \,.
\end{align}

For the theories considered here, Eq.~\eqref{Eq:DeltaEstimator} takes the form
\begin{multline}\label{Eq:DeltaEstimatorExplicit}
\DT(\hat{\theta}, z)
=  \frac{M'_{P}}{\H M_P} \left( \phi  -  v_{\|} + \int^{{\chi}}_0 d\Tilde{\chi}\,(\phi' + \psi') \right)+\\
-\frac{ M_{P, \varphi}}{M_P} \delta \varphi - \frac{ M_{P, X}}{M_P} \delta X \,,
\end{multline}
where only the explicit DE/MG-dependent effects are present.
In addition to the purely DE  contributions in the second line, only three effects contribute to $\DT$: the residual Doppler, SW and ISW effects.
Most importantly lensing convergence, which is the dominant contribution to GW- and   SN-radiation anisotropies, cancels out.
For particular classes of events, Eq.~\eqref{Eq:DeltaEstimator} could be directly evaluated for pairs of sources at the same position and redshift. In our analysis we require this to hold only statistically,  by integrating Eq.~\eqref{Eq:DeltaEstimator} over a joint redshift distribution and computing its angular power-spectrum:
\begin{align} \label{Eq:DTPowerSpectrum}
C_\ell^{\DT} = C_\ell^{\rm SN} + C_\ell^{\rm GW} - 2C_\ell^{\rm SN-GW}\,,
\end{align}
where $C_\ell^{\rm SN}$ ($C_\ell^{\rm GW}$) are the SN (GW) luminosity distance angular power-spectra, and $C_\ell^{\rm SN-GW}$ the cross-spectrum between the two.
In this form we need the redshift and position of GW/SN sources to be the same only on average, i.e. same redshift distributions and overlapping regions in the sky.

In Fig.~\ref{Fig:PowerSpectrumDelta} we show $C_\ell^{\DT}$ as a function of the source redshift  for the two representative DE models.
We consider the case of localized SN/GWs sources to study the redshift dependence of $C^{\DT}_\ell$.
In $f(R)$, the DE clustering component
is dominating the total angular power-spectrum, making its features manifest.
In the GBD model, instead,  the total signal is dominated by the Doppler effect.
Nevertheless, a detection of this signal still constitutes a direct  proof of the DE field's presence.

\section{Observational prospects}
We next investigate the detection prospects for the fluctuations of the GW luminosity distance via $C^{\rm GW}_\ell$, and DE clustering via $C^{\DT}_\ell$.
We consider the noise power-spectrum for both SN and GW, as given by only a shot-noise contribution~\cite{Cooray:2008qn,Calore:2020bpd}:
\begin{align} \label{Eq:SNnoisePowerSpectrum}
N^i_\ell = \frac{4\pi f_{\rm sky }}{N_i} \left(\frac{\sigma^i_{d_L}}{d^i_L} \right)^2
\equiv \frac{4\pi f_{\rm sky }}{N_i^{\rm eff}}
\,,
\end{align}
where $i = \{{\rm SN},{\rm GW}\}$ and $f_{\rm sky}$ is the sky fraction covered by observations, which we assume to be $f_{\rm sky}=1$ for simplicity.
We also define the effective number of sources, $N_i^{\rm eff}$, as the product of the number of events, $N_i$, in a given redshift bin and the ratio $\sigma^i_{d_L}/d^i_L$ related to the relative uncertainty on the luminosity distance which is proportional to the magnitude uncertainty. In this way $N_i^{\rm eff}$, which sets the overall noise levels, takes into account the number of events detected and the precision of each measurement.
As the signal decays in scale faster than $\propto \ell^{-2}$, we expect to have the best chance of measuring it
from large-scale observations. For this reason we assume that future localization uncertainties can be neglected~\cite{Baker:2019ync}.

The noise for the joint estimator of Eq.~\eqref{Eq:DTPowerSpectrum} is given by the sum of the two noise power-spectra for GW and SN, since we assume that any stochastic contribution is uncorrelated.
Consequently, the number of effective events needed for a detection of $C_\ell^{\DT}$ is given by the harmonic mean of the two single ones $N_{\DT}^{\rm eff} = \left((N^{\rm eff}_{\rm SN})^{-1}+ (N_{\rm GW}^{\rm eff})^{-1}\right )^{-1}$.
The error on a power-spectrum measurement is given by $\sigma(C_\ell) = \sqrt{2/(2\ell+1) f_{\rm sky}} [C_\ell + N_\ell]$,
and the corresponding signal-to-noise ratio is $(S/N)^2 = \sum_\ell (C_\ell / \sigma(C_\ell))^2$.
In the case of $C_\ell^{\rm GW}$ this applies directly, while for $C_\ell^{\Delta\varphi}$ one needs to do  full error propagation on Eq.~\eqref{Eq:DTPowerSpectrum}: the final result is the same, provided one uses for $N_{\DT}^{\rm eff}$  the harmonic mean given above.

The noise power-spectrum in Eq.~\eqref{Eq:SNnoisePowerSpectrum} is scale-independent so we can solve the inverse problem of determining the number of effective events needed to measure the power-spectra with a desired statistical significance.
In practice, we fix a target $S/N = 5$, and solve the equation of $S/N$ for $N^{\rm eff}$ both in the case of GW sources alone and $\DT$.

Finally, we investigate the scenario where the GW source redshift is unknown.
In this case we assume the shape of the GW redshift distribution as given in~\cite{Hirata:2010ba}, while the SN one as in~\cite{Hounsell:2017ejq}. Since the SN and GW redshift distributions need to match,
we take the product of the two and build the joint probability of measuring both SN and GW at the same redshift.

Intermediate cases in which the EM counterpart is not available, but estimates of the redshift distributions are obtained via statistical methods~\cite{MacLeod:2007jd,DelPozzo:2012zz}, would  fall in between the two extreme cases examined here.

\begin{table}[!ht]
\setlength{\tabcolsep}{8pt}
\centering
\begin{tabular}{@{}ll|ll|lll@{}}
\toprule
 & \multicolumn{1}{c}{GR} & \multicolumn{2}{c}{$f(R)$} & \multicolumn{2}{c}{GBD}  \\
\toprule
& $N_{\rm GW}^{\rm eff}$ &  $N_{\rm GW}^{\rm eff}$  &$N_{\DT}^{\rm eff}$ & $N_{\rm GW}^{\rm eff}$ & $N_{\DT}^{\rm eff}$ \\
\colrule
$z = 0.1$ & $10^{7}$ & $10^{7}$ &$10^{14}$& $10^{7}$ & $10^{12}$ \\
\colrule
$z = 0.3$ & $10^{8}$ & $10^{8}$ &$10^{15}$& $10^{8}$ & $10^{11}$\\
\colrule
$z = 0.7$ & $10^{8}$ & $10^{8}$ &$10^{16}$& $10^{8}$& $10^{12}$\\
\colrule
$z = 1.5$ & $10^{7}$ & $10^{7}$ &$10^{17}$& $10^{7}$& $10^{12}$ \\
\botrule
w/o $z$ & $10^{7}$ & $10^{7}$ &$10^{19}$& $10^{7}$ & $10^{14}$\\
\botrule
\end{tabular}
\caption{ \label{Table:NeffGW}
Effective number of events for a $5$-$\sigma$ detection of $C_\ell^{\rm GW}$ and $C_\ell^{\DT}$. }
\end{table}

Table~\ref{Table:NeffGW} summarizes the results reporting the number of effective sources for a $5\sigma$ detection of the angular power-spectra $C_\ell^{\rm GW}$ and $C_\ell^{\DT}$, both in the case of GW events with known as well as unknown redshifts (the latter designated as ``w/o $z$''). We also indicate the value of $N_{\rm GW}^{\rm eff}$ in GR,  for comparison. The detection threshold for GW luminosity distance fluctuations, $N^{\rm eff}_{\rm GW}$, does not change appreciably for the different scenarios, since we selected representative models sufficiently close to $\Lambda$CDM to satisfy current constraints. In fact, as shown in Fig.~\ref{Fig:PowerSpectrum}, $C_\ell^{\rm GW}$  is dominated by lensing convergence at high redshifts and by Doppler shift at low redshifts.
The former is {\it indirectly} modified by DE/MG, while the latter is also sensitive to the background configuration of the DE field: both  these effects are small in the considered models.
Since lensing convergence and Doppler effect dominate the angular correlations of GW sources,
it is not possible to distinguish the DE clustering contribution in $C_\ell^{\rm GW}$ within the total signal.

As far as $N_{\DT}^{\rm eff}$ is concerned, the results show that it is possible to detect the signal of the joint estimator in both cases of known and unknown redshifts.
In $f(R)$, this signal is dominated by the DE field fluctuations, as shown in Fig.~\ref{Fig:PowerSpectrumDelta}, hence allowing for its direct detection.
In the GBD model, the signal of the joint estimator is dominated by Doppler shift,  easier to detect, explaining the lower number of effective events compared to $f(R)$. In this case, one would not be able to distinguish directly the DE field inhomogeneities but its detection is still a  proof  of a time-dependent Planck mass.
Comparing the two scenarios of known and uknown GW source's redshift, we see that the number of effective events is larger in the latter case because a broader redshift range weakens the signal.
However, in this situation the events are not restricted to a redshift bin, hence  one can use the whole population of SN/GW sources provided that they are both present.
Nonetheless, the number of effective events required  is very high, suggesting that the detection precision per source has to  improve   to eventually measure such signal.
In fact, we remark that $N_i^{\rm eff}$ is the effective number of sources, \emph{the real number of events can be lowered by having smaller statistical errors on the single detection}.
As an example, in order to measure the DE signal, the detection of a population of about $10^6$
GW sources and about the same number of SN events in a redshift bin at $z>1$, would require a precision, per event, of about $\sigma_{d_L}/d_L \sim 10^{-6}$ in the case of $f(R)$, and $\sim 10^{-3}$ for the GBD model. Since the required effective number of events scales quadratically with per-event precision, $\sigma_{d_L}/d_L$, but only linearly with number of events, increasing precision is likely a better strategy.

\section{Discussion and Outlook}
Fluctuations in the DE field can distinctively alter the propagation of GWs with respect to light.
In this work, we derived the expression for such effects in a class of DHOST theories, generalizing the results of~\cite{Garoffolo:2019mna}.  And, by combining luminosity distance measurements from GW and SN sources, we built an estimator for the \emph{direct} detection of the imprint of the DE fluctuations, that does not rely on non-gravitational interactions between DE and known particles.
This signal cannot be mimicked by other effects and, as such, it provides a distinctive evidence for DE/MG.

Even if the DE clustering signal is below cosmic variance, any detection of our joint estimator would be a convincing proof of a running Planck mass, as we showed for two specific models. Reversely, it can be used to place complementary bounds on theories of dynamical dark energy non-minimally coupled to gravity, along similar lines of recent forecasts as in~\cite{Belgacem:2019pkk,Lagos:2019kds} for the case of standard sirens.
Since we exploit angular correlations at large scales, we expect our method not to be affected by screening mechanisms nearby sources.

On the other hand, one should leverage as much as possible on the precision of the measurement; for instance, given the number of SN/GW events (of order $10^6$,  at least in the higher redshift bins) that can be observed with future SN surveys~\cite{Ivezic:2008fe,Abell:2009aa} and space-based interferometers~\cite{phynney1,Kawamura:2011zz},  a detection would be possible, if one decreases the statistical error on each measure according to table \ref{Table:NeffGW}.
Notice also that for our estimates we considered an ideal case: the number of events needed for a detection might be higher to deal with possible systematic effects.
This suggests that future facilities might have to develop new technologies and observational strategies to meet these detection goals.
We leave it to future work to determine whether a detection of the signal we propose can be aided
by studying additional MG models, synergies with large scale structure surveys or considering different sources of GW/EM signals. For example, future experiments will detect large numbers of binary white dwarfs (BWD)~\cite{McNeill:2019rct} on galactic scales and much beyond~\cite{Maselli:2019mzt,Kinugawa:2019uey}. BWD are supposed to be progenitors of Type-Ia SN in the so-called double degenerate scenario~\cite{Maoz:2011iv}, offering a common source for  GW and SN signals (see e.g.~\cite{Gupta:2019okl}). In this case,  Eq.~\eqref{Eq:DeltaEstimator} holds locally and  $\DT$  could be directly reconstructed in configuration space, provided that non-linearities and MG screening effects can be properly taken into account.
It will also be interesting to use our general formula, Eq.~\eqref{Eq:DeltaGW}, to investigate whether other DE cosmological models based on DHOST (see e.g.~\cite{Crisostomi:2017pjs,Belgacem:2019pkk,Arai:2019zul}) lead to signals easier to detected with fewer sources.

%%%%%%%%%%%%%%%%%

\appendix

\section{Derivation of Equation~(1)}\label{App1}
The DHOST action we consider is \cite{Crisostomi:2019yfo}
\begin{equation}\label{dhostac1}
S=\int d^4 x \sqrt{-g}\left[ K+ G\,\Box \varphi+F R+\frac{3\,F_{,X}^2}{2\,F}\,
\varphi_{;\mu}\,\varphi^{;\mu\sigma}\,\varphi_{;\sigma \nu}\,\varphi^{;\nu}
\right]\,,
\end{equation}
where $K$, $G$, $F$ are functions of $\varphi$ and $X\,=\,-\varphi_{,\mu} \varphi^{,\mu}/2$ and  $; \mu$ indicates a covariant derivative along coordinate $x^\mu$.

We split the metric and the scalar field as a slowly-varying  plus a high-frequency small fluctuations:
\begin{eqnarray}
     &g_{\mu\nu} = \bar g_{\mu\nu} + h_{\mu\nu} \qquad &\mbox{with} \qquad |h_{\mu\nu}| \ll |\bar{g}_{\mu\nu}|\,, \label{MetricSplit} \\
     &\varphi = \bar \varphi + \varphi_r \qquad &\mbox{with} \qquad |\varphi_r| \ll |\bar \varphi|\,. \label{ScalarSplit}
\end{eqnarray}
If $L$ is the typical lengthscale over which the background varies, i.e. $|\partial \bar g, \partial \bar \varphi | \sim 1/L$, and $\lambda$ the one of the high-frequency modes, i.e. $|\partial h, \partial \varphi_r  |\sim 1/ \lambda$, then $L \gg \lambda$. We introduce the small parameter $\epsilon \equiv \lambda / L \ll 1$ controlling the expansion in derivatives of the high-frequency perturbations (see e.g.~\cite{Maggiore:1900zz, Isaacson:1967zz}).

The perturbations $\{h_{\mu\nu}, \varphi_r\}$ transform under  diffeomorphism in the standard way.
We assume  the hierarchy
$|\varphi_r | \sim \epsilon |h_{\mu\nu}|$ between
  the high-frequency scalar  and metric fluctuations.
   Since a gauge transformation mixes $\varphi_r$ and $h_{\mu\nu}$, the latter assumption guarantees that Eq.~\eqref{MetricSplit} holds after the gauge transformation. We use the gauge freedom to choose
\begin{eqnarray}
  \bar \nabla^\mu \hat h_{\mu\nu} &=& 0 \, \qquad \mbox{and}\qquad \varphi_r = 0\,, \label{GaugeChoice}
\end{eqnarray}
where $\hat h_{\mu\nu} = h_{\mu\nu} -\frac12 \bar g_{\mu\nu} (\bar g^{\alpha\beta} h_{\alpha\beta} )$.
As shown in \cite{Garoffolo:2019mna}, the two conditions in Eq.~\eqref{GaugeChoice} are compatible only when $\varphi_r$ propagates at the same speed of the tensor modes.
This condition can be imposed by choosing suitable relations between the functions $F, K$ and $G$ evaluated on the slowly varying background $\{ \bar g_{\mu\nu}, \bar \varphi \}$.

The system of linearized  equations of motion
is organized in powers of $\epsilon$: second derivatives acting on $h_{\mu\nu}$ are the leading order contributions since $|\partial \partial h| \sim h / \epsilon^2$, while those with first derivatives constitute the next-to-leading order since $|\partial h| \sim h / \epsilon$ and $\epsilon \ll 1$. Terms not containing  derivatives of  high-frequency modes can be consistently discarded.

We use the geometric optics ansatz,
\begin{equation} \label{GOanstaz}
   \hat h_{\mu\nu} = {\cal A}_{\mu\nu} e^{i \theta / \epsilon} = {\cal A} {\bf e}_{\mu\nu} e^{i \theta / \epsilon}\,,
\end{equation}
where ${\cal A}$ is the amplitude, ${\bf e}_{\mu\nu}$ the polarization tensor and $\theta$ the phase of the gravitational wave (GW).
Note that the gauge choice made ensures that the waves are transverse but not traceless.
Hence, we decompose ${\cal A}_{\mu\nu}$ as
\begin{equation}
  {\cal A}_{\mu\nu} = {\cal A}^{\rm GW}_{\mu\nu} + {\cal A}^S_{\mu\nu} = {\cal A}^{\rm GW} ({\bf e}^{+}_{\mu\nu} + {\bf e}^{\times}_{\mu\nu}) + {\cal A}^S {\bf e}^S_{\mu\nu}\,,
\end{equation}
where we assumed that the $+$ and $\times$ polarizations have the same amplitude. The wave vector is defined as $k_\rho \equiv - \bar \nabla_\rho \theta$ and, for theories such as~\eqref{dhostac1}, high-frequency perturbations propagate at the speed of light thus $k^\mu k_\mu = 0 $ and $k^\mu \bar \nabla_\mu k^\nu = 0$. The polarization tensors are such that $k^\mu {\bf e}^{+}_{\mu\nu} = k^\mu {\bf e}^{\times}_{\mu\nu} = k^\mu {\bf e}^{S}_{\mu\nu} = 0$ and $\bar g^{\mu\nu} \,{\bf e}^{+}_{\mu\nu} = \bar g^{\mu\nu} \, {\bf e}^{+}_{\mu\nu} = 0$.
Following the  procedure of in \cite{Garoffolo:2019mna},  the evolution equation of the amplitude of the tensor modes is
\begin{equation} \label{EqAmplitudeGeneral}
\bar \nabla_\rho \left( k^\rho \left({\cal A}^{\rm GW} \right)^2\right)\,=\,-k^\rho \, \bar \nabla_\rho\,\ln{\bar F}\,\left({\cal A}^{\rm GW} \right)^2\,,
\end{equation}
where $\bar F = F|_{\bar \varphi}$.

In order to derive Eq.~\eqref{Eq:DeltaGW} we use the \textit{Cosmic Rulers} formalism (see e.g.  \cite{Bertacca:2017vod,Schmidt:2012ne}) which allows to compute the effects of large scale structures (LSS) on the propagating GW. In particular, the {\it observer-frame} is used as reference system to compute the relevant physical quantities related to GWs. The latter frame is different from the {\it real-frame} because, when observing, we use coordinates that flatten the past light-cone of the GW.

In order to study the propagation of GW through cosmic inhomogeneities we choose for $\{\bar g_{\mu\nu}, \bar \varphi \}$
\begin{eqnarray}
     d \bar{s}^2 &=&  a^2(\eta)[ - (1+2\phi(x)) d\eta^2 + (1 - 2\psi(x)) d \bar{x}^2]\,, \label{MetricPoisson}\\
     \bar \varphi &=& \varphi_0(\eta) + \delta \varphi(x)\,, \label{ScalarField}
\end{eqnarray}
where $\phi, \psi$ are the two gravitational potentials in Poisson's gauge and $\delta \varphi$ is the DE field fluctuation.

Eq.~\eqref{EqAmplitudeGeneral} is similar to the one in~\cite{Garoffolo:2019mna} but in this case of $\bar F = \bar F[\bar \varphi, \bar X]$ instead of $\bar F = \bar F[\bar \varphi]$.
Hence we outline the steps of the procedure and report the explicit computation only of the new contribution. The full derivation can be found in~\cite{Garoffolo:2019mna}.

Firstly, we perform a conformal transformation and define $ \hat{g}_{\mu\nu} \equiv \Bar{g}_{\mu\nu} / a^2$ and $\hat{k}^\mu \equiv k^\mu  a^2$
and, using $ \text d / \text d \chi \equiv \hat k^\mu \hat \nabla_\mu $, the evolution equation becomes
\begin{eqnarray}
  \frac{\text{d}}{ \text{d} \chi}\ln ({\cal A}^{\rm GW}\,a) &=& - \frac{1}{2} \bigg( \hat{\nabla}_\rho \hat{k}^\rho + \frac{\text{d}}{ \text{d} \chi} \ln \bar F \bigg) \, , \label{EQAmplitudeGWConformal}
\end{eqnarray}
where $\hat \nabla_\mu $ is the covariant derivative with respect to $\hat g_{\mu\nu}$.

After the conformal transformation,  $\hat{g}_{\mu\nu} = \eta_{\mu\nu} + \delta  \hat g_{\mu\nu}$, and observer-frame and real-frame would coincide if $\delta  \hat g_{\mu\nu} = 0$.
However, in the linear regime, $\delta  \hat g_{\mu\nu}$ is small, thus observed and real quantities will differ by a small amount. This fact can be exploited to build a map between the two frames valid at linear order in $\delta \hat g_{\mu \nu}$, where observed-quantities are identified as the $0^{th}$ order terms of the expansion and real-quantities will be given by the sum of the relative observed-quantity plus a small correction.
Considering, as an example, the amplitude of the GW, ${\cal A}^{\rm GW}$, this will be expanded as ${\cal A}^{\rm GW} = {\cal A}^{\rm GW}_o + \Delta {\cal A}^{\rm GW}$, where ${\cal A}^{\rm GW}_o$ is the observed amplitude of the incoming GW which satisfies Eq.~\eqref{EQAmplitudeGWConformal} with $\delta  \hat g_{\mu\nu} = 0$.
Hence, ${\cal A}^{\rm GW}_o$ is given by
\begin{eqnarray}
  \frac{\text d}{\text d \bar{\chi}} \ln ({\cal A}_0^{\rm GW}\Bar{a}  \Bar{\chi} \sqrt{F_0}) &=& 0\,,
\end{eqnarray}
where $F_0 = \bar F|_{\varphi_0}$ and whose solution is
\begin{equation}
  {\cal A}_0^{\rm GW}(\bar{x}^0,\bar{\chi}) = \frac{\mathcal{Q}^{\rm GW }}{\Bar{a}(\bar{x}^0)\Bar{\chi}\sqrt{F_0}} =  \frac{\mathcal{Q}^{\rm GW }\,(1+z)^2}{\Bar{d}_L\sqrt{F_0}} \label{AmplitudeGWZero}\,.
\end{equation}
 In the latter equation $\mathcal{Q}^{\rm GW }$ is the integration constant which depends on the production mechanism of the GW, $\bar \chi$ is the observed comoving distance, $\Bar{a} = (1+z)^{-1}$ is the observed scale factor and $\bar{d}_L = (1+z)\bar{\chi}$ is the observed average luminosity distance taken over all the sources with the same observed redshift. From Eq.~\eqref{AmplitudeGWZero} one can define the ${\rm GW }$ luminosity distance as
$\bar{d}^{\rm GW}_L = \bar{d}_L\sqrt{F_0}$.

The perturbation $\Delta {\cal A}^{\rm GW}$ is the solution of the first order (in  $\delta \hat g_{\mu\nu}$) amplitude's evolution equation, namely

\begin{eqnarray}
  \left[\frac{\text{d}}{ \text{d} \chi}\ln ( {\cal A}^{\rm GW} a) \right]^{(1)}  &=& - \frac{1}{2} \left[\hat{\nabla}_\rho \hat{k}^\rho \right]^{(1)} -\frac12 \left[\frac{\text{d}}{ \text{d} \chi}\ln \bar F \right]^{(1)}  \,\,.
\end{eqnarray}
The computation of the three terms not explicitly dependent on the DE field is present in literature (see e.g.~\cite{Bertacca:2017vod}). We illustrate the computation of the novel contribution, as in this work $\bar F =  F [\bar \varphi, \bar X]$, while in~\cite{Garoffolo:2019mna} it was $\bar F =  F [\bar \varphi]$.
In particular, this is
\begin{align}
&\left[\frac{\text{d}}{ \text{d} \chi} \ln \bar F (x(\chi)) \right ]^{(1)} = \nonumber\\
&\qquad=\bigg(1-\frac{\text d \delta\chi}{\text d \bar{\chi}} \bigg)\frac{\text d}{\text d \Bar{\chi}} \ln \left[ F_0(\bar x(\bar \chi)) \left(1 + \frac{\Delta F (\bar x(\bar \chi))}{F_0 (\bar x(\bar \chi))} \right) \right] = \nonumber\\
 &\qquad= \delta \chi \frac{\text d^2 \ln F_0 (\bar x(\bar \chi))}{\text d \bar \chi^2} + \frac{  \text d  (\delta x^\mu\, \Bar{\partial}_\mu \ln F_0(\bar x(\bar \chi)))}{\text d \Bar{\chi}} + \nonumber\\
 &\qquad\quad+ \frac{\text d}{ \text d \bar \chi} \left( \frac{\delta F (\bar x(\bar \chi))}{F_0(\bar x(\bar \chi))} \right)\,, \label{Four}
\end{align}
where we have used:   the relation between real, $\chi$, and observed, $\bar \chi$, affine parameters: $\chi = \bar \chi + \delta \chi$; the relation between real, $x^\mu (\chi)$, and observed, $ \bar x^\mu (\bar \chi)$, GW geodesics: $x^\mu(\chi) = \bar x^\mu (\bar \chi)+  \delta x^\mu (\bar{\chi}) + \bar{k}^\mu  \delta \chi $; and the expansion at linear order $\bar F(x(\chi)) = F_0(\bar x (\bar \chi)) + [\delta F (\bar x(\bar \chi))+ (\delta x(\bar{\chi}) + \bar{k}^\mu  \delta \chi) \bar \partial_\mu F_0(\bar x(\bar \chi))]$
where $\bar \partial_\mu = \partial / \partial \bar x^\mu$.
Combining all of the four contributions together we find
\begin{align}
\Delta \ln &{\cal A}^{\rm GW} =  \,\kappa + (\phi+\psi) - \frac{1}{{\bar \chi}}\int^{{\bar \chi}}_0 d\Tilde{\chi}\,(\phi + \psi) \nonumber\\
& -\phi \bigg(\frac{1}{{\bar \chi}\H} -\frac{F'_)}{2\H F_0}\bigg)- v_{\|} \bigg(1-\frac{1}{{\bar \chi}\H} +\frac{F'_0}{2 \H F_0} \bigg)   \nonumber\\
	&  + \left(\,1  - \frac{1}{{\bar \chi}\H} + \frac{F'_0}{2\H F_0}\right) \int^{{\bar \chi}}_0 d\Tilde{\chi}\,(\phi' + \psi')  - \frac{\delta F}{2F_0}\,, \label{DeltaDL}
\end{align}
where ${\cal H} \equiv \bar a' / \bar a$, primes are derivative with respect to conformal time, $v_{\|}$ is the component of the peculiar velocity field along the line of sight and $\kappa$ is the standard weak lensing convergence factor. In the main text $M_P \equiv \sqrt{F}$, so $F'_0 / 2 F_0 = (M'_P/ M_P)|_{\varphi_0}$ and
\begin{equation}
  \frac{\delta F}{2F_0} =  \frac{M_{P,\varphi}}{M_P} \,\delta \varphi + \frac{M_{P,X}}{M_P} \, \delta X\,,
\end{equation}
where $M_{P,\varphi}$ and $ M_{P,X}$ are the derivatives of $M_P$ with respect to its arguments.
The amplitude in the real-frame is given by
\begin{align}
    {\cal A}^{\rm GW}(\chi) &= {\cal A}_0^{\rm GW}(1+\Delta \ln {\cal A}^{\rm GW}) = \frac{\mathcal{Q}^{\rm GW}\,(1+z)^2}{\Bar{d}^{\rm GW}_L (1- \Delta \ln {\cal A}^{\rm GW})} \, , \label{AmplitudeGW}
\end{align}
so that the luminosity distance fluctuation, Eq.~\eqref{Eq:DeltaGW} of the main text, is
\begin{eqnarray}
  \frac{\Delta d^{\rm GW}_L}{\Bar{d}^{\rm GW}_L} &=& \frac{d^{\rm GW}_L - \Bar{d}^{\rm GW}_L}{\Bar{d}^{\rm GW}_L} = - \Delta \ln {\cal A}^{\rm GW}(\bar{\chi})\,, \label{DeltaDLGW}
\end{eqnarray}
where $\Delta \ln {\cal A}^{\rm GW}(\bar{\chi})$ is given in Eq.~\eqref{DeltaDL}.

\section{Derivation of Equation~(5)}\label{App2}
In scalar-tensor theories of gravity, such as Eq.~\eqref{dhostac1}, photons do not directly couple to the extra scalar field. Hence the amplitude of the EM radiation,  named ${\cal A}^{\rm SN}$, satisfies
\begin{eqnarray}
  \bar \nabla_\rho \left( k^\rho \left({\cal A}^{\rm SN}\right)^2\right)\,&=&\,0\,, \label{EQAmplitudeSN}
\end{eqnarray}
where the covariant derivative is associated with  $\bar g_{\mu\nu}$.
Eq.~\eqref{EQAmplitudeSN} is formally equal to the evolution equation of the amplitude of GW in General Relativity hence the Cosmic Rulers formalism gives the same results as in~\cite{Bertacca:2017vod}:
\begin{align}
\frac{\Delta d^{\rm SN}_L}{\Bar{d}^{\rm SN}_L} =&  -\kappa + (\phi+\psi) + \frac{1}{{\bar \chi}}\int^{{\bar \chi}}_0 d\Tilde{\chi}\,(\phi + \psi)  +\frac{\phi}{{\bar \chi}\H}- \nonumber\\
&\, - v_{\|} \bigg(1-\frac{1}{{\bar \chi}\H} \bigg) - \left(\,1  - \frac{1}{{\bar \chi}\H}\right) \int^{{\bar \chi}}_0 d\Tilde{\chi}\,(\phi' + \psi') \,.
\label{DeltaDLSN}
\end{align}
Combining Eq.~\eqref{DeltaDLSN} and Eq.~\eqref{DeltaDLGW} one gets Eq.~\eqref{Eq:DeltaEstimator}.

\smallskip

\section{Code implementation}\label{App3}
We rewrite the $(k, \ell)$ multipoles of $ \Delta d^{\rm GW}_L/\Bar{d}^{\rm GW}_L$, Eq.~(3) of the main text,  in  a notation which is more suitable for direct implementation in \texttt{EFTCAMB}~\cite{Hu:2014oga}. The resulting expression can be written in terms of the different sources,  highlighting each relativistic or modified gravity effect,

\begin{align}\label{DeltaDLGW_multipole}
\left(\frac{\Delta d^{\rm GW}_L}{\bar{d}^{\rm GW}_L}\right)^{W}_{k \, \ell} & =\int^{\eta_A}_0 \text{d} \eta \, j_\ell(k \chi) \bigg \{ S^{\rm GW}_{\kappa} + S^{\rm GW}_{vol}  + S^{\rm GW}_{Sh}+ \nonumber \\
&\:+  S^{\rm GW}_{SW}  +S^{\rm GW}_{Dop} + S^{\rm GW}_{ISW}+ S^{\rm GW}_{\delta \varphi} \bigg\}\,,
\end{align}
with $\eta_A$ is the conformal time corresponding to $z = +\infty$ and
\begin{eqnarray} \label{SourcesGW}
S^{\rm GW}_\kappa(\eta) &=& (\phi_{{k}}+ \psi_{{k}})  \int^{\eta}_0 \text{d} \tilde{\eta}\,\frac{\ell(\ell+1)}{2}  \frac{(\Tilde{\chi}-\chi)}{\Tilde{\chi} \chi} W(\tilde{\eta}) \nonumber\\
S^{\rm GW}_{vol}(\eta) &=& -W(\eta) (\phi_{{k}}+\psi_{{k}} )\,,\nonumber\\
S^{\rm GW}_{Sh}(\eta)&=&(\phi_{{k}} + \psi_{{k}})  \int^{\eta}_0 \text{d} \tilde{\eta}\,  \frac{W(\tilde{\eta})}{\Tilde{\chi}} \nonumber \\
S^{\rm GW}_{SW}(\eta) &=& W(\eta)\bigg(\frac{1}{\chi\H} -\frac{M'_P}{\H M_P}\bigg) \psi_{{k}}\,,\nonumber \\
S^{\rm GW}_{Dop}(\eta) &=& - \partial_\eta \left[W(\eta) \bigg(1-\frac{1}{\H\chi} +\frac{M'_P}{\H M_P} \bigg) v \right] \nonumber\\
S^{\rm GW}_{ISW}(\eta) &=& (\phi'_{{k}} + \psi'_{{k}}) \int^{\eta}_0 \text{d} \tilde{\eta}\, W(\tilde{\eta}) \left(\,1 + \frac{M'_P}{\H M_P} - \frac{1}{\chi\H}\right)\,,\nonumber\\
S^{\rm GW}_{\delta \varphi}(\eta) &=& W(\eta) \bigg(\frac{M_{P,\varphi}}{M_P} \,\delta \varphi + \frac{M_{P,X}}{M_P} \, \delta X\bigg)\,,
\end{eqnarray}
where $W(\eta) = (1+z) \H W(z)$.
In \eqref{SourcesGW} we assumed $v^i(\bar{k},\eta) = i  k^i\,v(\eta)$, namely that the peculiar velocity field is irrotational.
Similarly, for the joint estimator $\Delta_\varphi$ we find
\begin{eqnarray}
S^{\Delta_\varphi}_{SW}(\eta) &=& W(\eta)\frac{M'_P}{\H M_P}\psi \nonumber \\
S^{\Delta_\varphi}_{Dop}(\eta) &=&  \partial_\eta \left[W (\eta) \frac{M'_P}{\H M_P}v\right] \nonumber\\
S^{\Delta_\varphi}_{ISW}(\eta) &=& -\bigg(\phi'_{{k}} + \psi'_{{k}}\bigg) \int^{\eta}_0 \text{d} \tilde{\eta}\, W(\tilde{\eta}) \left(\,\frac{M'_P}{\H M_P}\right)\,,\nonumber\\
S^{\Delta_\varphi}_{\delta \varphi}(\eta) &=& - W(\eta)  \bigg(\frac{M_{P,\varphi}}{M_P} \,\delta \varphi + \frac{M_{P,X}}{M_P} \, \delta X\bigg)\,.
\end{eqnarray}

\section{Comparison with other works in literature}
In this section we make a comparison between our results and those of the two works~\cite{Dalang:2019rke, Dalang:2020eaj}. The main difference between the two approaches is the gauge choice imposed on the high-frequency modes  after the  split of the metric and scalar field as in~\eqref{MetricSplit} and~\eqref{ScalarSplit}.
The two fields rapid perturbations transform under a gauge transformation as
\begin{eqnarray}
  h'_{\mu\nu} &=& h_{\mu\nu}  - \bar \nabla_\mu \xi_\nu -  \bar \nabla_\nu \xi_\mu   \,,\\
  \varphi'_r &=& \varphi_r - \xi^\mu \,\bar \nabla_\mu \bar \varphi \,,
\end{eqnarray}
hence a gauge  transformation mixes $h_{\mu\nu}$ and $\varphi_r$, and it might invalidate the split between background  and perturbed metric, namely eq.~\eqref{MetricSplit}. For this reason, in~\cite{Garoffolo:2019mna} it is assumed that $|\varphi_r| \sim \epsilon |h_{\mu\nu}|$, so that eq.~\eqref{MetricSplit} holds even after a gauge transformation. \cite{Garoffolo:2019mna} opts for the gauge conditions $\varphi_r = 0$ and
$\bar \nabla^\nu \hat h_{\mu\nu} =0$, where $\hat h_{\mu\nu}$ is the trace-reversed metric perturbation.
These two gauge choices are compatible under the assumption that the high-frequency scalar and tensor modes propagate at the same speed, namely the one of light. This assumption is at the basis of ~\cite{Garoffolo:2019mna}, alternative to the gauge choice of the recent ~\cite{Dalang:2020eaj} (while in~\cite{Dalang:2019rke} the high frequency scalar excitation, $\varphi_r$, is not considered).
In our approach, we ensure that $\varphi_r$ propagates at the same speed of $h_{\mu\nu}$ imposing a suitable condition on the functions $F, K$ and $G$, evaluated on the slowly varying background, while~\cite{Dalang:2020eaj} opts for choosing $K_{XX} = 0$ and $G_X = 0$.
Despite these differences in the approach, we find the same evolution equation for the amplitude of the tensor modes ${\cal A}^{\rm GW}$, namely eq. (4.14) in~\cite{Garoffolo:2019mna}, eq. (54) in~\cite{Dalang:2019rke} and eq. (95) in~\cite{Dalang:2020eaj}. This is the quantity we are interested in and focus on in the present work.
%In fact, the two methodologies lead to different results only regarding the polarization of the GWs as reflection of the different gauge choices made. However,
The dynamics of ${\cal A}^{\rm GW}$ in the three works is the same since, in  theories in which $\{ h_{\mu\nu}, \varphi_r\}$ both propagate at the same speed, the evolution equation of ${\cal A}^{\rm GW}$ decouples form the one of ${\cal A}^{S}$.

%%%%%%%%%%%%%%%%%%

\begin{acknowledgments}
We thank Alvise Raccanelli for helpful discussions.
%
%CC thanks Enzo Branchini for very useful discussions.
%
MR is supported in part by NASA ATP Grant No. NNH17ZDA001N, and by funds provided by the Center for Particle Cosmology.
AG and AS acknowledge support from the NWO and the Dutch Ministry of Education,
Culture and Science (OCW) (through NWO VIDI Grant No. 2019/ENW/00678104 and from the D-ITP consortium).
GT is partially funded by STFC grant ST/P00055X/1.
DB and SM acknowledge partial financial support by ASI Grant No. 2016-24-H.0.
Computing resources were provided by the University of Chicago Research Computing Center through the Kavli Institute for Cosmological Physics at the University of Chicago.
\end{acknowledgments}

\bibliographystyle{apsrev4-1}
\bibliography{references}

\end{document}